\documentclass[twocolumn,showpacs,preprintnumbers,amsmath,amssymb]{revtex4}

\usepackage[dvips]{graphicx}
\usepackage{dcolumn}
\usepackage{bm}

\begin{document}
\def\bfB{\mbox{\bf B}}
\def\bfQ{\mbox{\bf Q}}
\def\bfD{\mbox{\bf D}}
\def\etal{\mbox{\it et al}}

\title{From reversing to hemispherical dynamos}

\author{Basile Gallet, Fran\c{c}ois P\'etr\'elis}

\address{Laboratoire de Physique Statistique de l'Ecole Normale
Sup\'erieure, UMR CNRS 8550, 24 Rue Lhomond, 75231 Paris Cedex 05, France.}

\date{\today}

\begin{abstract}
We show that hemispherical dynamos can result from weak equatorial symmetry breaking of the flow in the interior of planets and stars. Using a model of spherical dynamo, we observe that the  interaction between a dipolar and a quadrupolar mode can localize the magnetic field in only one hemisphere  when the equatorial symmetry is broken. This process is shown to be related to the one that is responsible for reversals of the magnetic field. These seemingly very different behaviors are thus understood in a unified framework.
\end{abstract}


\maketitle

Planets and stars seem to be able to generate hemispherical dynamos.  During the Maunder minimum, the number of  spots at the Sun's surface decreased and the sunspots were mostly localized in the southern hemisphere \cite{soleil}. It is  believed that the magnetic field was then also dominant in this hemisphere \cite{weiss}.  Although Mars does not act as a dynamo  anymore, it has been suggested that it did in the past and created an hemispherical  magnetic field  \cite{mars}. In addition to these  observations, direct numerical simulations  have clearly identified hemispherical dynamos  \cite{busse}. Another surprising behavior is that of the Earth's magnetic field which  reverses randomly in time. Reversals have  been observed in various numerical simulations \cite{revue} and in a laboratory experiment \cite{vks}. Among the numerical  studies, events have also been reported  where the magnetic field is much larger in one hemisphere \cite{glatz}. Reversing and hemispherically localized  magnetic fields are two seemingly very different features  of astrophysical, numerical and experimental dynamos.

We explain these two different behaviors in a unified framework: they both result from the breaking of the  equatorial symmetry of the flow.  As direct numerical simulations have already exhibited the two behaviors at stake, we focus on a simple kinematic $\alpha^2$ dynamo model in which equatorial symmetry can be broken. 
The obtained results are then generalized in the framework of dynamical systems theory:  when the equatorial symmetry is weakly broken, reversals or hemispherical localization are shown to be two generic behaviors of any dynamo which two most unstable modes are of dipolar and quadrupolar symmetries.

We consider a sphere of radius $R$ that contains a fluid  of electrical  conductivity $\sigma$ and magnetic permeability equals to that of vacuum $\mu_0$. The outside medium is vacuum. The inductive processes of the flow are modelled by $\alpha$ effects localized on two shells of vanishingly small radial thickness located at radius $\chi R$ and $\xi R$. On the outer shell, an $\alpha$ effect converts the poloidal magnetic field into  toroidal field. This latter is converted back into poloidal field  on the inner shell. We thus write the $\alpha$ tensor in spherical coordinates $(r,\theta,\phi)$ as 
\begin{equation}
\alpha_{i j}=\alpha_{\theta \theta} \delta(r-\xi R) \delta_{i \theta} \delta_{j \theta}+\alpha_{\phi \phi}  \delta(r-\chi R)  \delta_{i \phi} \delta_{j\phi}\,.
\end{equation}
The two scalars $\alpha_{\theta \theta}$ and $\alpha_{\phi \phi}$ are supposed to depend on $\theta$ only and are written 
\begin{eqnarray}
\alpha_{\phi \phi} & = a_1+b_1 \cos(\theta)+c_1 \cos(2 \theta) + d_1 \cos(3 \theta)\,,\nonumber\\
\alpha_{\theta \theta} & =  a_2+b_2 \cos(\theta)+c_2 \cos(2 \theta) + d_2 \cos(3 \theta)\,.
\label{defcoeff}
\end{eqnarray}
Odd dependences in latitude model an equatorially symmetric flow  whereas even ones break this symmetry. We solve the induction equation for this $\alpha^2$ dynamo and look for axisymmetric  eigenmodes ${\bf B(r)} e^{\tilde{p} t}$ solutions of 
\begin{equation}
\tilde{p} \, {\bf B}=\nabla \times (\alpha  {\bf B})+(\mu_0 \sigma)^{-1}\nabla^2 {\bf B}\,,
\end{equation}
for $r\le R$ and that  match a potential field at $r = R$.  Similar calculations were performed for  $\alpha-\omega$ dynamos with  equatorially symmetric effects only \cite{Deinzler,Schmitt}.   
From now on, we use the dimensionless growth rate $p=\tilde{p}\mu_0\sigma R^2$ and $\alpha$ effects $Ra_1=\mu_0 \sigma a_1$, $Rb_1=\mu_0 \sigma b_1$... Moreover we set the position of the shells $\chi=0.55$ and $\xi=0.9$.

\begin{figure}[htb!]
\centerline{\includegraphics[width=6cm]{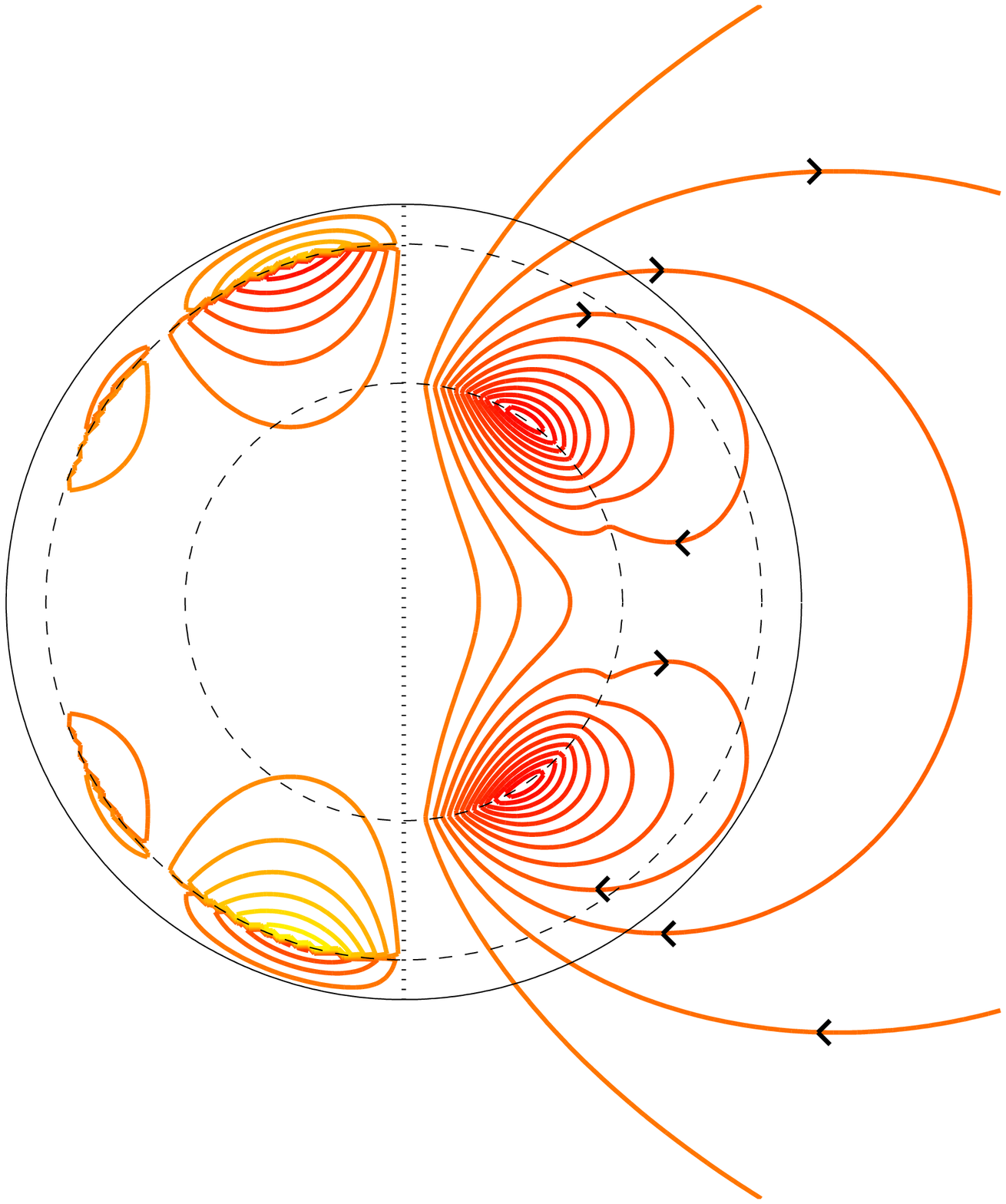}}\centerline{\includegraphics[width=6cm]{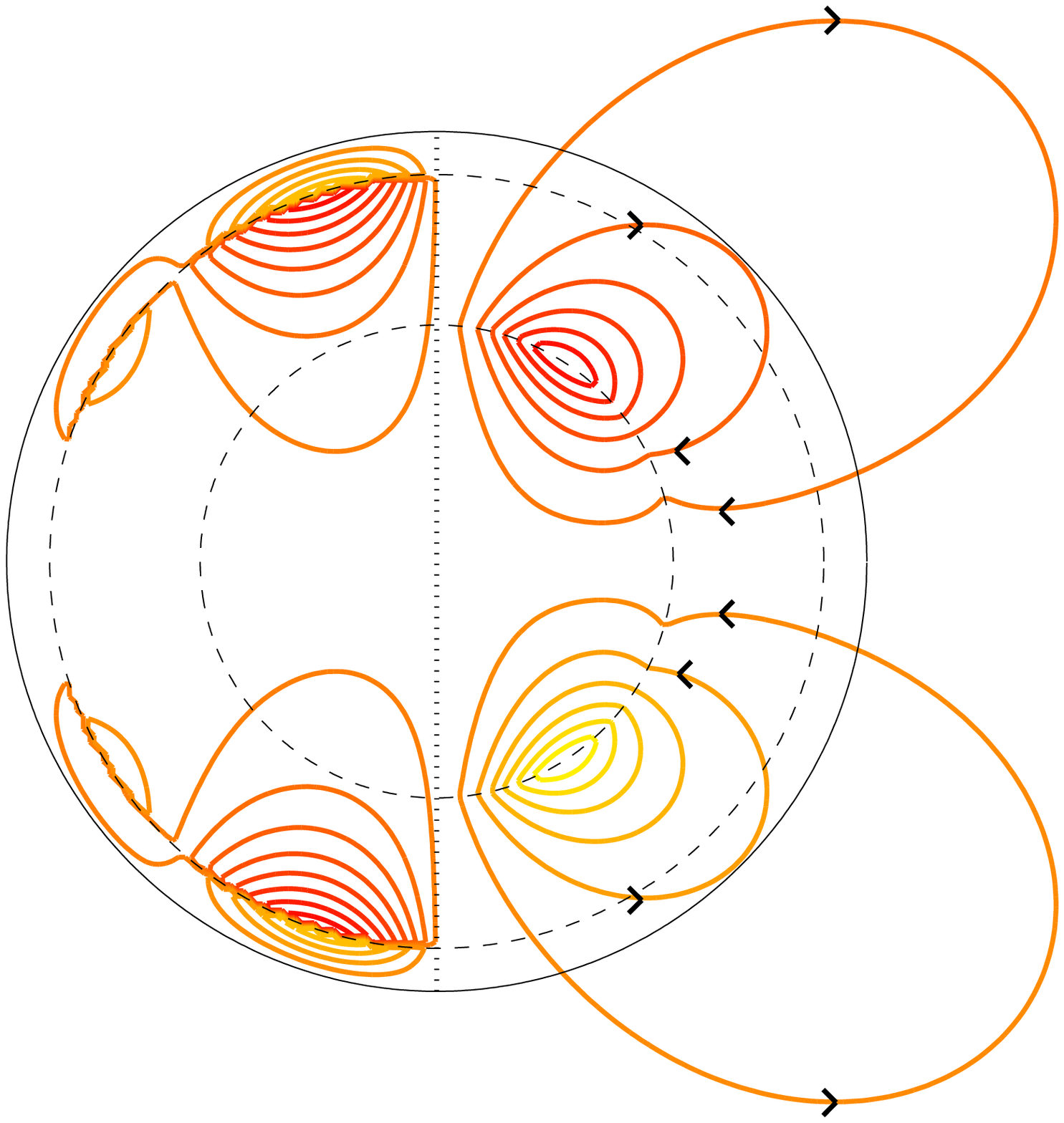}}
\vspace{-.75cm}
\caption{Unstable magnetic field modes when the flow properties are equatorially symmetric (see parameters in the text). The dashed circles are the two shells where the $\alpha$ effects are located. On the right hand side of the revolution axis (dotted), the lines of poloidal field are drawn. Lines of constant toroidal fields are shown on the left hand side.  Top (resp. bottom):  mode of dipolar (resp. quadrupolar) symmetry. } 
\label{fig1}
\vspace{-.3cm}
\end{figure}

We first consider an equatorially symmetric configuration with $Rb_1=Rb_2=20$, $Rd_2=25$, all other parameters being  zero.  Two modes are then linearly unstable. The most unstable eigenmode is of dipolar symmetry and is represented in fig. \ref{fig1} (top). Its growth rate is $p \simeq 1.26$.   For this set of parameters, a second mode is unstable and its  growth rate is only slightly smaller $p \simeq 1.21$. This mode is represented in fig. \ref{fig1} (bottom). It  is even under equatorial reflection and is  therefore of quadrupolar symmetry.  
The existence of two unstable modes depends on the values of the parameters. We observe in general that the dipolar and quadrupolar modes have similar growth rates when the $\alpha$ effects are larger close to the poles. Roughly speaking, both hemispheres generate a magnetic structure. These structures are far from each other and thus only weakly coupled: the situation with poloidal fields of same polarity, {\it i.e.} a dipole, does not strongly differ from a situation with opposite poloidal fields, {\it i.e.} a quadrupole. 
This argument, see for instance \cite{Moffat}, explains why a dipole and a quadrupole have similar growth rates and are likely to be unstable for close values of the parameters.

\begin{figure*}
\centerline
{\includegraphics[width=4.5cm]{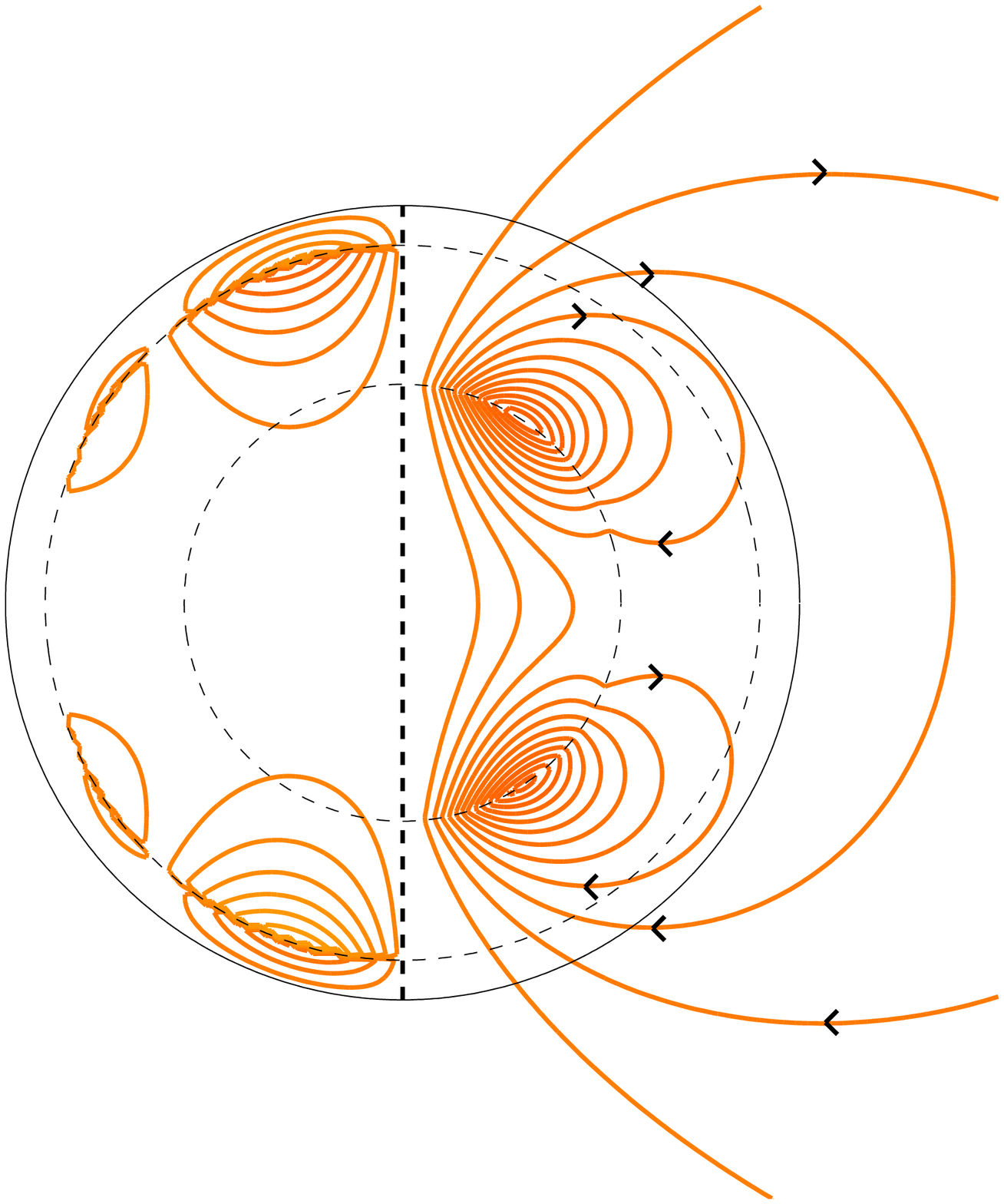}\ \includegraphics[width=4.5cm]{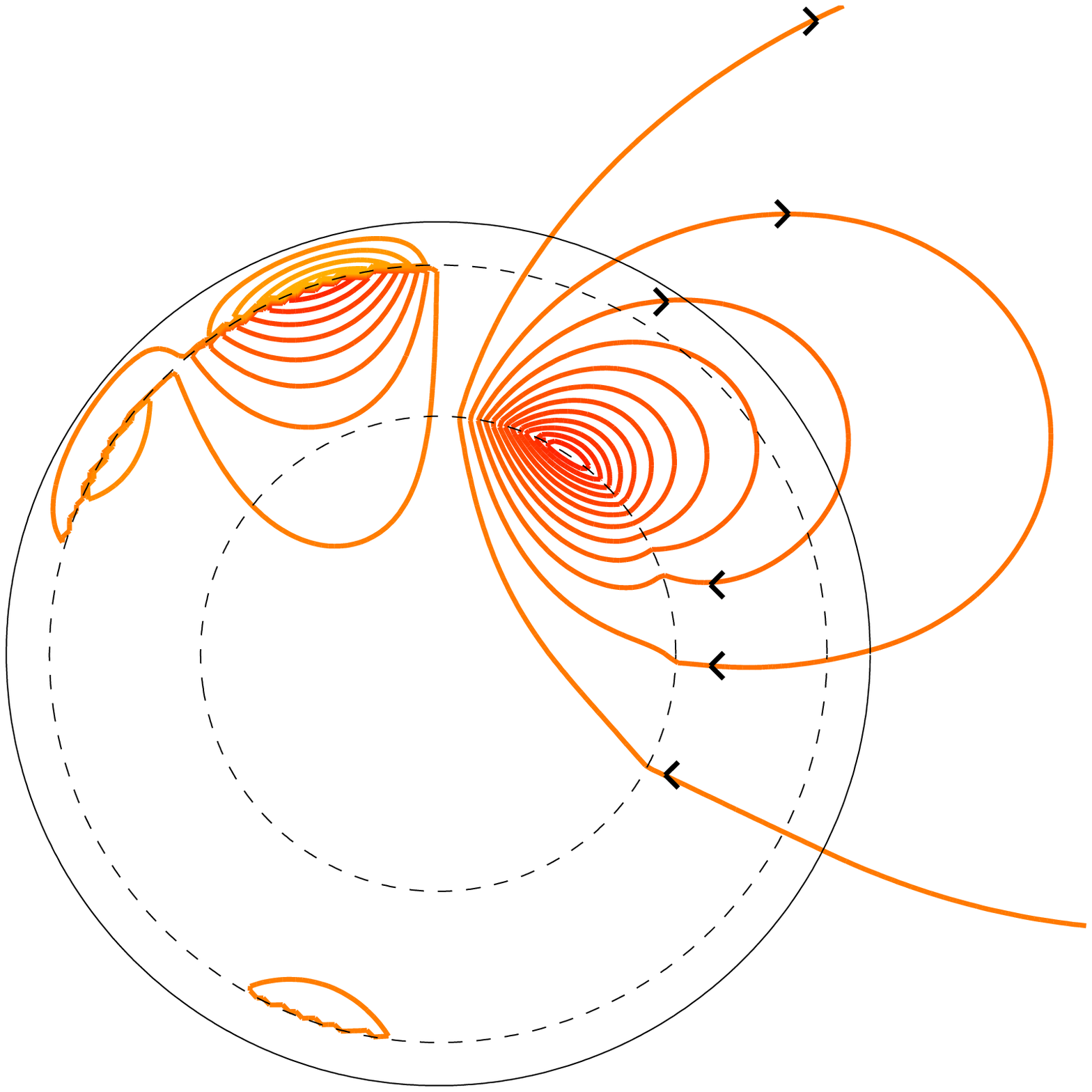}\
\includegraphics[width=4.5cm]{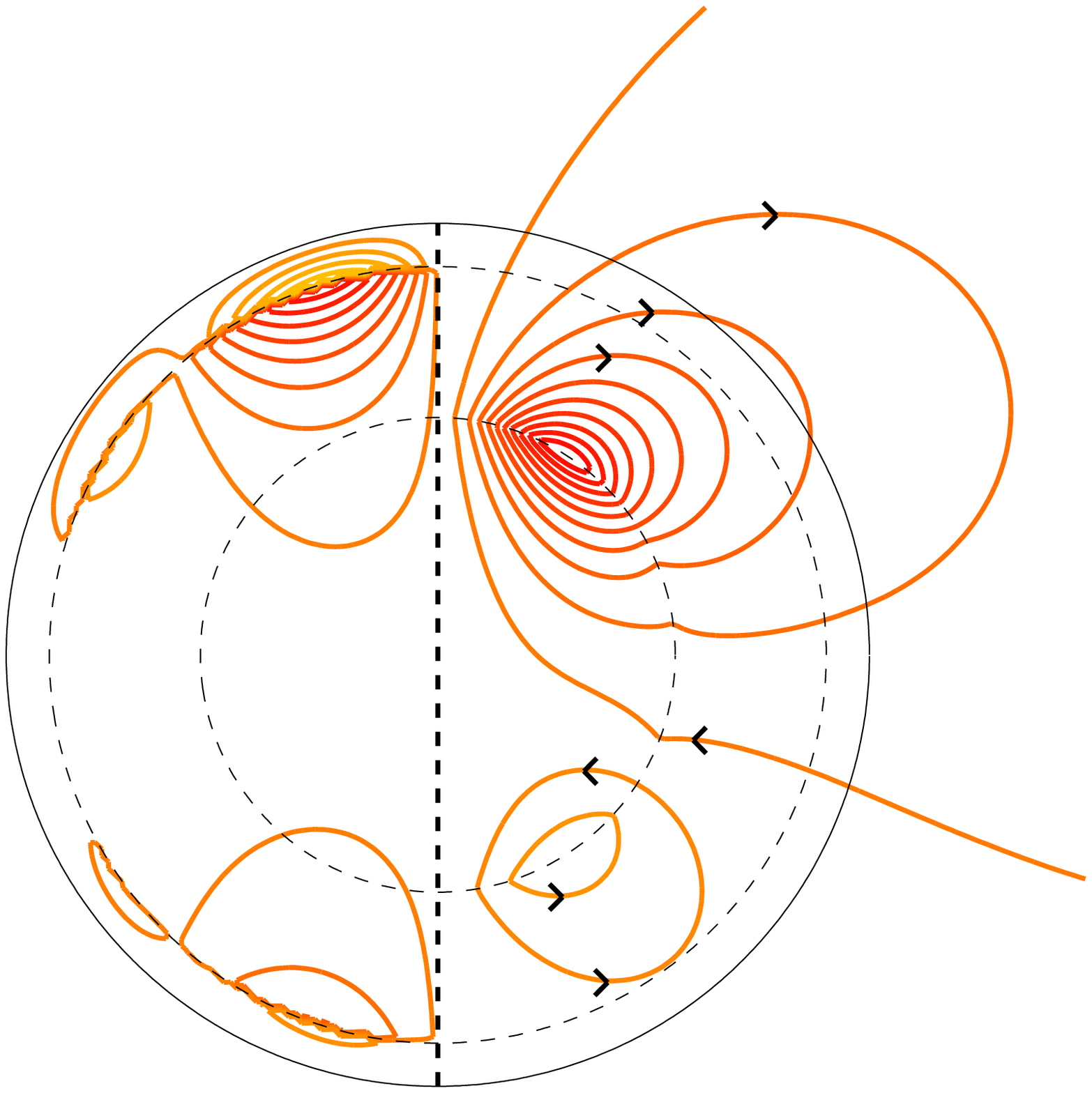}}
\centerline
{\includegraphics[width=4.5cm]{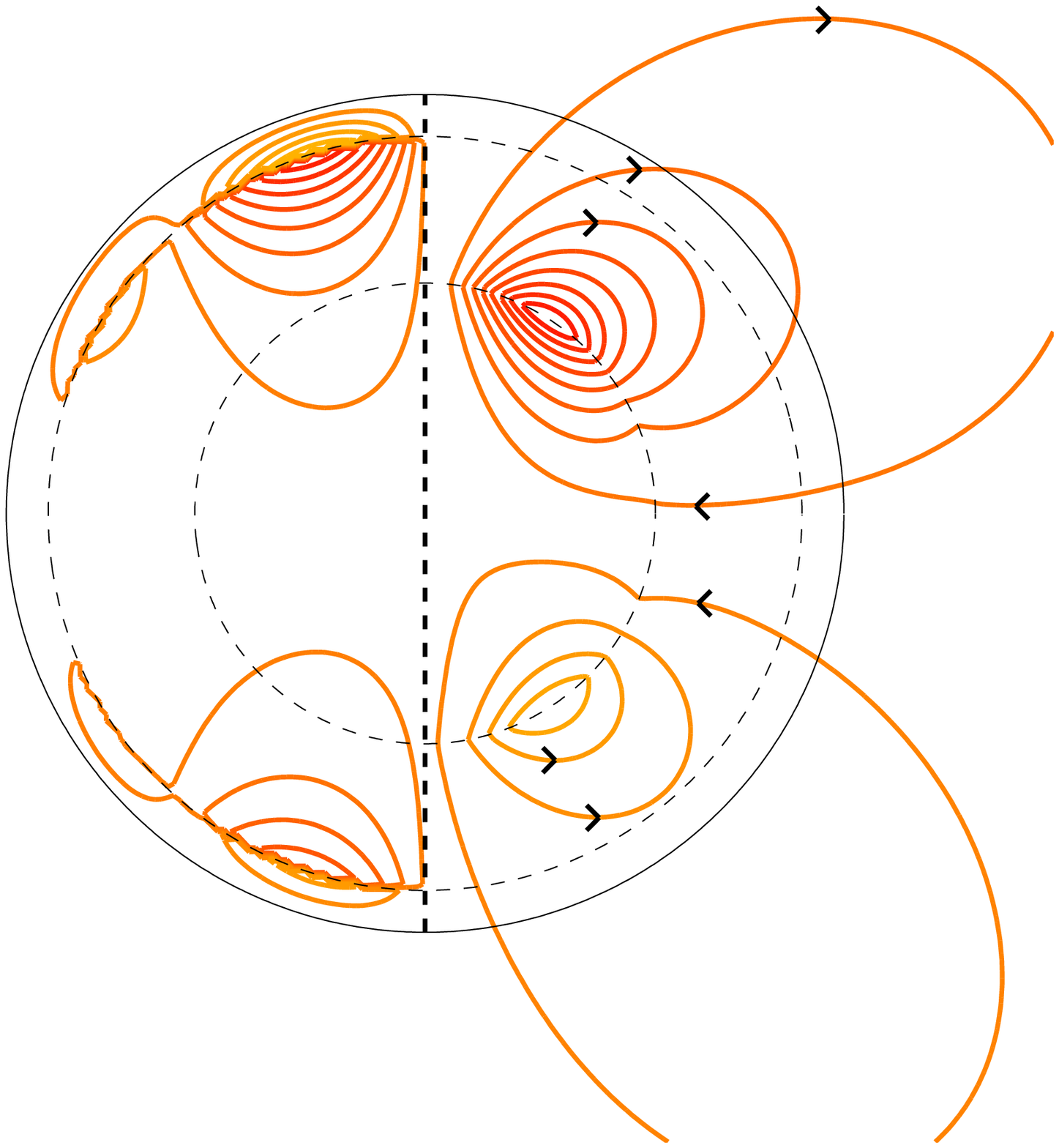}\ \includegraphics[width=4.5cm]{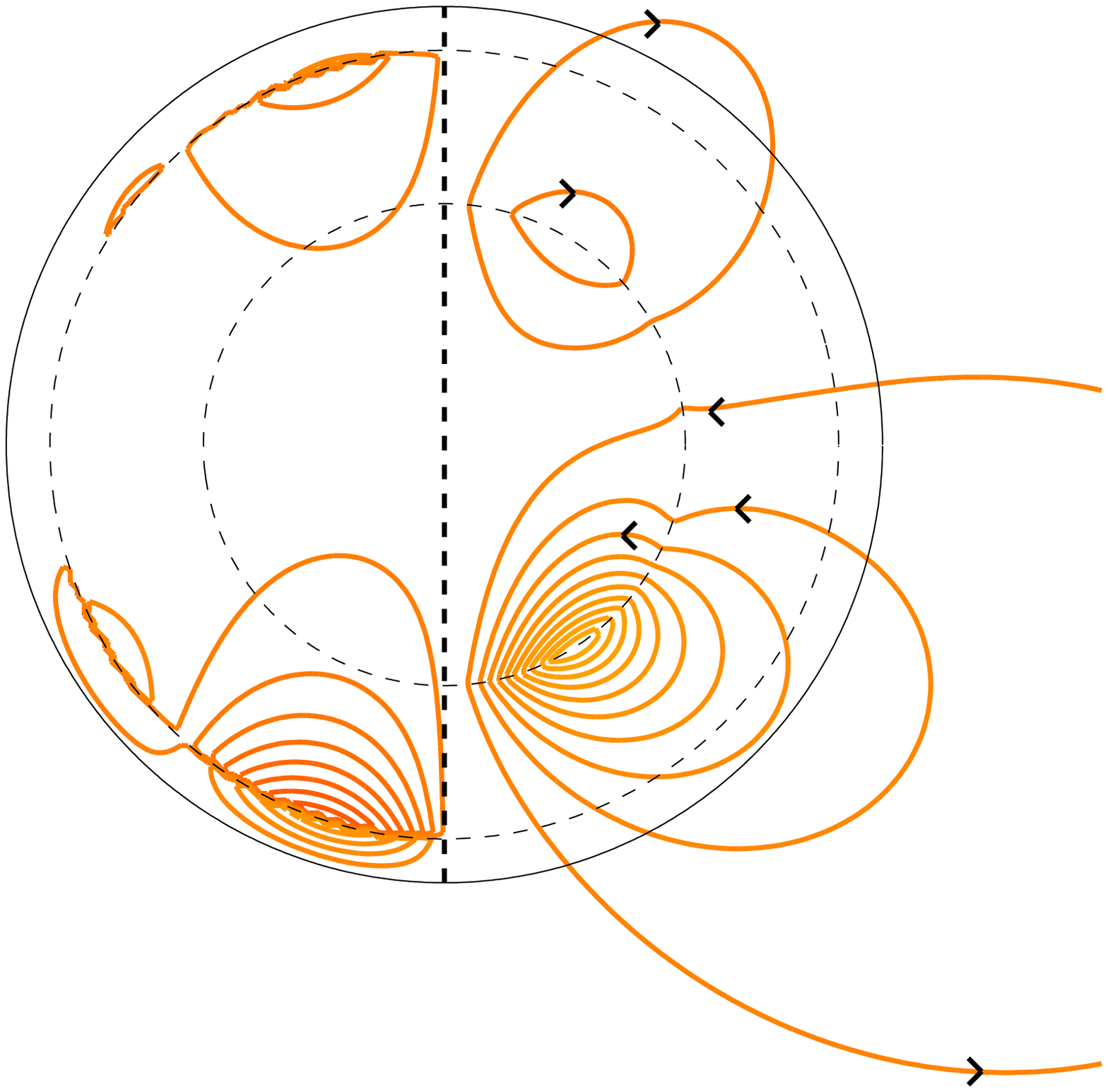}\
\includegraphics[width=4.5cm]{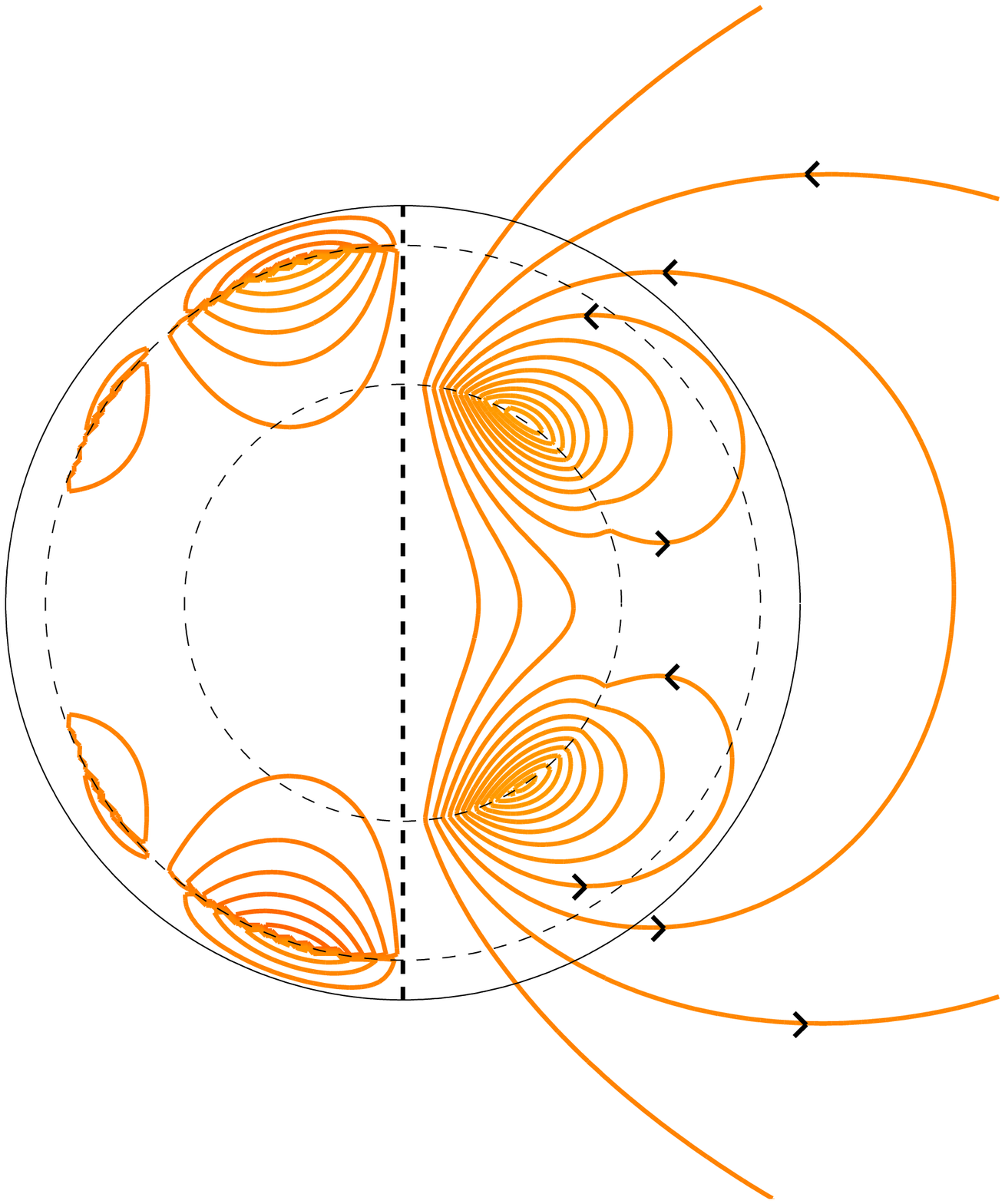}}
\vspace{-.5cm}
\caption{Evolution of the magnetic field during a reversal obtained when the equatorial symmetry is weakly broken. From left to right and top to bottom,  half a cycle is represented at  time $t=0$ (arbitrary), $t=T/10$,  $2 T/10$, $3 T/10$, $4.5 T/10$ and $T/2$, $T$ being the period. The dipolar field evolves toward a quadrupole  before a dipole of opposite polarity is recovered.}
\label{fig2}
\vspace{-.5cm}
\end{figure*}

Starting from this equatorially symmetric configuration, we now break equatorial symmetry by adding an $\alpha$ effect which has an even dependence on latitude. A first class of equatorial symmetry breaking leads to a decrease in the difference between the growth rates of the two modes, $\Delta p$. To illustrate this regime, we consider $Rc_1=-Rc_2=P$ where $P$ measures the symmetry breaking and $Ra_1=Ra_2=0$. When $P$ increases, $\Delta p$ decreases but  both modes  remain stationary until  a bifurcation occurs at $P_c=0.78$ where the eigenvalues of the two most unstable modes are equal. For $P$ slightly larger than $P_c$, they are complex conjugates. This corresponds to an exponentially growing oscillatory magnetic field.  A snapshot of the time-evolution of the field during half a period is shown in fig. \ref{fig2} for $P=1$. The dipole  reverses and during a reversal, a quadrupolar structure is observed. The main effect of nonlinearities is to saturate the amplitude of the magnetic field (see discussion below): we have removed the exponential growth in this representation.  We point out that the transition from stationary to oscillatory  dynamo occurs for small symmetry breaking. For $P=1$, the symmetry breaking effects ($Rc_1$ and $Rc_2$) are of the order of $5 \%$ of the equatorially symmetric terms.

It has already been observed that oscillatory dynamos can be obtained from  stationary modes when parameters are changed. Numerical studies of $\alpha-\omega$  dynamos display transitions from stationary to oscillatory decaying eigenmodes that most likely result from an interaction between two modes \cite{roberts}.  
Competition between dipolar modes only has been studied for $\alpha^2$ dynamos \cite{stefani}: for particular forms of $\alpha$ effect, {\it i.e.} diagonal and dependent on $r$ only, oscillatory modes can be observed. Here, we have shown that very weak  equatorial symmetry breaking is sufficient to  couple a dipolar and a quadrupolar mode so that a time periodic regime is obtained. When nonlinear terms are taken into account, this transition occurs through  a saddle-node bifurcation and  below the onset of bifurcation, turbulent fluctuations of the flow can trigger random reversals \cite{FP08}. Such time dependent fluctuations are not taken into account in our calculation. Nevertheless, in the regime of random reversals, the snapshots obtained during a reversal would be similar to that displayed in fig. \ref{fig2}: once a reversal has been initiated, its evolution is mainly deterministic and corresponds to half a period of the nonlinear oscillation obtained above the saddle-node bifurcation. 

\begin{figure}
\centerline{\includegraphics[width=6cm]{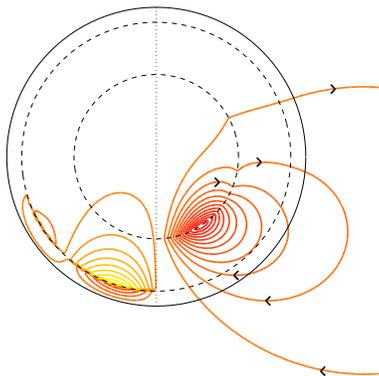}}
\vspace{-.5cm}
\caption{Most unstable magnetic mode for a symmetry breaking that leads to field localization (see parameters in the text).} 
\label{fig3}
\vspace{-.25cm}
\end{figure}

\begin{figure}
\centerline{\includegraphics[width=6.5cm]{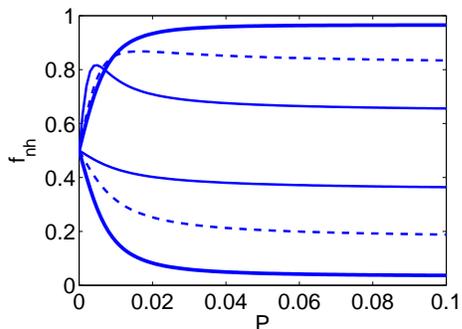}}
\vspace{-.5cm}
\caption{Relative energy at the surface of the northern hemisphere $f_{nh}=E_n/(E_n+E_s)$  as a function of the intensity $P$ of the equatorial symmetry breaking effect. The curves correspond to  (thick) $r=R$, (dashed) $r=2R$ and (solid) $r=5R$ for the two unstable eigenmodes, the  lower curve corresponding to the most unstable eigenmode.} 
\label{fig4}
\vspace{-.25cm}
\end{figure}

We have also identified other  terms that break equatorial symmetry but do not lead to reversals:  the two most unstable modes remain stationary and the difference between their growth rates $\Delta p$ increases. These  modes display a surprising spatial structure: the field is almost completely localized in only one hemisphere. This is illustrated with the set of parameters $Ra_1=Ra_2=-P$ and $Rc_1=Rc_2=0$ and  in fig. \ref{fig3}  the most unstable mode is represented for $P=0.2$.  Note that the second most unstable mode is localized in the other hemisphere. 
To quantify this localization, we compute $E_n$  (resp. $E_s$) the energy of the field measured at the surface of the northern (resp. southern) hemisphere for a fixed radius $r$. We plot in fig. \ref{fig4}  the relative energy in the northern hemisphere $f_{nh}=E_n/(E_n+E_s)$ as a function of $P$.   For large $P$, the ratio $f_{nh}$ saturates to a  limit value $f^{\infty}$. This value depends on $r$ and tends to $1/2$ when $r$ becomes very large. Indeed, for  large $r/R$, the field is dominated by the dipole for which $f_{nh}=0.5$. Nevertheless, we point out that even for $r=2 R$, $83 \%$ of the energy is still localized in the southern hemisphere  in the case shown  in fig. \ref{fig3}.  This effect can thus be observed at the surface of planets and stars even if dynamo action takes place much deeper in the fluid core. At first sight, one might believe that the localization is a straightforward consequence of the $\alpha$-effect being more intense in one hemisphere. We present here the simplest  form of symmetry breaking but we have obtained the same behavior with different ones, such as $Ra_1=-Ra_2=-P$: the two $\alpha$ effects are enhanced in different hemispheres but hemispherical localization still occurs. Even more surprisingly, we point out that a very weak symmetry breaking has a drastic effect on the field localization. In  fig. \ref{fig4}, for $P\ge 0.05$ {\it i.e.} as soon as $Ra_1/Rb_1=0.25 \%$, more than $95 \%$ of the energy is located in the south hemisphere at $r=R$. 

The fact that a symmetry breaking of a few percents leads to such a strong modification of the geometry of the eigenmode is a consequence of the interaction between two modes of instability. Indeed,  hemispherical localization can be understood as a competition between a dipolar mode, ${\bf D (r)}$ with amplitude $d(t)$ and a quadrupolar mode ${\bf Q (r)}$ with amplitude $q(t)$ from which the total magnetic field is determined, ${\bf B}= d(t) {\bf D (r)}+ q(t) {\bf Q (r)}$. The amplitude equation at linear order for $A=d+i\, q$ is  
\begin{equation}
\dot{A}=(\mu_r+i\, \mu_i) A+(\nu_r+i\, \nu_i) \bar{A}\,,
\label{eqA}
\end{equation}
where the coefficients $\mu_r$, $\mu_i$, $\nu_r$ and  $\nu_i$ are real. 
Searching for a phase equation, we write $A=r \exp{(i \Theta)}$, and obtain 
\begin{equation}
\dot{\Theta}=\mu_i+\nu_i \cos{(2 \Theta)}-\nu_r \sin{(2 \Theta)}\,.
\label{eqt}
\end{equation}
When the flow is equatorially symmetric, $\mu_i=\nu_i=0$ and $\nu_r$ measures the difference of growth rates between the two modes. When equatorial symmetry is broken, $\mu_i$ and $\nu_i$ increase. If $\mu_i$ increases faster than $\nu_i$, a regime of reversals is obtained \cite{FP08}. 
 
Our results show that some terms that break equatorial symmetry do  not lead to reversals, {\it i.e.}   lead to  $\nu_i$  increasing faster than  $\mu_i$. This is responsible for hemispherical localization of the field. Indeed, if $\nu_i$ is large compared to $\mu_i$ and $\nu_r$, the solutions of (\ref{eqt}) remain stationary and are $\Theta\simeq \pi/4\,, 3 \pi/4\, [\pi]$. These values correspond to a total field ${\bf B}\propto {\bf D} \pm {\bf Q}$. Its spatial structure is  easily understood: in one hemisphere the amplitudes of the dipolar and  quadrupolar modes are of same sign and the field is strong, whereas in the other hemisphere, the dipole and quadrupole are roughly opposite and their sum results in a field of very low intensity. 
 We have highlighted the fact that, for hemispherical localization to occur,  symmetry breaking $\alpha$ effects do not need to be large.  
This traces back to  the similar growth rates of the dipolar and  quadrupolar modes. More precisely, the symmetry breaking effects, of order $\nu_i$, need only to be larger than the growth rate difference $\nu_r$. Both terms can thus remain small.   

The role of nonlinear terms can be understood in this low dimensional approach. Terms in higher powers of $A$ and $\bar{A}$ have to be taken into account in (\ref{eqA}). Their main effect is to saturate the field amplitude $r$. They also modify the coefficients of (\ref{eqt}), but  the qualitative features of reversals are the same as those displayed in fig. \ref{fig2}. More precisely, nonlinear terms would change the critical value $P_c$ and the precise temporal dynamics of $\Theta$. However,  these two effects are of higher order in the field amplitude and remain small provided the modes are only slightly supercritical. This is also true for  hemispherical dynamos.

Thanks to the low dimensional approach we have shown that the results from the $\alpha^2$ model are generic and can be used to describe the dynamics observed in astrophysical, numerical and experimental  dynamos.  
Direct numerical simulations of convective dynamos have displayed 
hemispherically localized fields where "both quadrupolar and dipolar fields contribute nearly equal magnetic energy"  even though "the convection flow has lost little of its equatorial symmetry in the presence of the hemispherical magnetic field" \cite{busse}.  Numerical models of Mars have also obtained hemispherical dynamos \cite{mars}. Our results give a simple explanation for these localized fields: if the dipolar and quadrupolar modes have similar growth rates, even a very weak  equatorial symmetry breaking is enough to localize the field. As far as reversals are concerned, a numerical study of the Von Karman Sodium experiment displayed a  transition from steady to oscillatory magnetic field resulting from the coupling between a dipolar and a quadrupolar mode \cite{Gissinger}.

Studies aimed at modelling the Maunder minimum considered an oscillatory dipole in competition with  an oscillatory quadrupole  with similar pulsations \cite{knobloch}:  nonlinear coupling terms between these modes can generate hemispherically localized magnetic fields that remain time-periodic. The pure dipolar state is still a solution but is unstable towards a mixed polarity state. Further above onset, chaotic behaviors have also been observed \cite{knobloch,brooke}.
Our  mechanism for localization also applies to this situation and does not require nonlinear effects. 
In contrast to the aforementionned scenario, it  does not result from a secondary instability of a pure dipolar mode but from the modification of the dipolar solution induced by the breaking of  the flow's equatorial symmetry.

We can speculate on the origin of the symmetry breaking. First the flows are turbulent and velocity fluctuations are expected to break equatorial symmetry. Second, the boundary conditions can be asymmetric: the pattern of  thermal flux at the core mantle boundary has been shown to influence the frequency of reversals \cite{flux}.

Our simplified model captures some  widespread features of  dynamos. Since our results are constrained by symmetry considerations,  more realistic models  should display the same behaviors. In other words, all dynamos that have two unstable modes of dipolar and quadrupolar symmetry  behave the following way. Breaking the flow's equatorial symmetry leads to two different regimes:  a first class of symmetry breaking, associated to large $\mu_i$ in eq. (\ref{eqt}), results in reversing and oscillating dynamos. A second class, associated to large $\nu_i$, generates hemispherical dynamos. The fact that these apparently very different regimes result from the same flow feature is  appealing and  strengthens  the importance of the competition between dipolar and quadrupolar modes. In particular, we expect that several new observations of   hemispherical magnetic fields will be reported in   numerical, experimental or astrophysical dynamos.

Out of the dynamo context, our study shows that the competition between two modes of different symmetries leads to rich behaviors. It is known that reversals can be observed in large scale flows generated over a turbulent background  in thermal convection or in periodically driven flows \cite{kolmo}. We expect that a weak symmetry breaking of the forcing can also generate a localization of these large scale flows. This could  strongly affect the mixing properties or the thermal transfer efficiency of these systems.



\vspace{-.75cm}
\begin{thebibliography}{99}
\vspace{-2cm}\bibitem{soleil} J.C. Ribes and E. Nesme-Ribes, Astron. \& Astrophys. {\bf 276}, 549 (1993).

\bibitem{weiss} J. H. Thomas and N. O. Weiss, {\it Sunspots and Starspots}, Cambridge University Press, Cambridge (2008).

\bibitem{mars} S. Stanley et al., Science {\bf 321}, 1822 (2008).

\bibitem{busse} E. Grote and F. H. Busse, Phys. Rev. E {\bf 62}, 4457 (2000).

\bibitem{revue} See for instance, P. H. Roberts and G. A. Glatzmaier,  Rev. Mod. Phys. {\bf 72}, 1081 (2000). 

\bibitem{vks} M. Berhanu et al., Europhys. Lett. {\bf 77}, 59001 (2007). 

\bibitem{glatz} G. A. Glatzmaier and P. H. Roberts, Nature {\bf 377}, 203 (1995). 



 
\bibitem{Deinzler} W. Deinzler and M. Stix, Astron. \& Astrophys. {\bf 12}, 111 (1971).

\bibitem{Schmitt} D. Schmitt, Astron. \& Astrophys. {\bf 174}, 281 (1987).

\bibitem{Moffat} H. K. Moffatt, {\it Magnetic Field Generation in Electrically  
Conducting Fluids}, Cambridge University Press, Cambridge (1978).



\bibitem{roberts} P. H. Roberts, Phil. Trans. Roy. Soc. A {\bf 272}, 663 (1972),  see for instance fig. 1 p 674.

\bibitem{stefani} F. Stefani and G. Gerbeth, Phys. Rev. Lett. {\bf 94}, 184506 (2004).


\bibitem{FP08}  F. P\'etr\'elis  and  S. Fauve, J. Phys. Condens. Matter {\bf 20}, 494203 (2008).  F. P\'etr\'elis et al., Phys. Rev. Lett.  {\bf 102}, 144503 (2009). 

\bibitem{Gissinger} C. Gissinger,  arXiv:0906.3792v1 (2009). C. Gissinger et al., arXiv:0904.3343v1 (2009).


\bibitem{knobloch} E. Knobloch and A. S. Landsberg, Mon. Not. R. Astron. Soc. {\bf 278}, 294 (1996).

\bibitem{brooke} J.M. Brooke et al., Astron. \& Astrophys. {\bf 332}, 339 (1998).

\bibitem{flux} G. A. Glatzmaier et al., Nature {\bf 401}, 885 (1999). 


\bibitem{kolmo} R. Krishnamurti and  L. N. Howard, Proc. Natl. Acad. Sci. {\bf 78}, 1981 (1981). B. Liu and J. Zhang, Phys. Rev. Letters, {\bf 100}, 244501 (2008). J. Sommeria, J. Fluid Mech. {\bf 170}, 139 (1986).


\end{thebibliography}
\end{document}